\documentclass[11pt]{article}
\usepackage{latexsym}
\usepackage{epsfig}

\def\bg#1{\mbox{\boldmath$#1$}}

\newcommand{\anabla}{{\overrightarrow{\nabla}}\!\!\!\!\!\!{\overleftarrow{\nabla}}}

\newcommand{\del}{\partial}
\newcommand{\beq}{\begin{eqnarray}}
\newcommand{\eeq}{\end{eqnarray}}
\newcommand{\be}{\begin{eqnarray*}}
\newcommand{\ee}{\end{eqnarray*}}
\newcommand{\bk}{{\bf k}}
\newcommand{\bp}{{\bf p}}
\newcommand{\bq}{{\bf q}}
\newcommand{\br}{{\bf r}}

\newcommand{\ra}{\rightarrow}
\newcommand{\e}{\epsilon}

\newcommand{\nn}{\nonumber}

\begin{document}

\centerline{\Large\bf {Effective-Range Corrections to the Proton-Proton Fusion Rate}}
\vskip 10mm
\centerline{Xinwei Kong$^1$ and Finn Ravndal\footnote{On leave of absence from Institute 
            of Physics, University of Oslo, N-0316 Oslo, Norway}} 
\medskip
\centerline{\it Department of Physics and Institute of Nuclear Theory,}
\centerline{\it University of Washington, Seattle, WA 98195, U.S.A}

\bigskip
\vskip 5mm
{\bf Abstract:} {\small Proton-proton fusion is considered in the effective field theory
of Kaplan, Savage and Wise. Coulomb effects are included systematically in a non-perturbative
way. Including the dimension-eight derivative coupling which determines the effective ranges
of the scattering amplitudes, next-to-leading order corrections to the fusion rate are
calculated. When the renormalization mass is well above the characteristic energies of the
system, this contribution gives a rate which is eight percent below the standard 
value. The difference can be due to an unknown counterterm which comes in at this order.}

During the last year much progress has been made in understanding the low-energy 
properties of few-nucleon systems from 
the effective field theory of Kaplan, Savage and Wise\cite{KSW_1}. Both elastic and
inelastic processes can be considered. Bound states like the deuteron can be treated via
its interpolating field which replaces the need for explicit wavefunctions\cite{KSW_2}. 
A more general review of the whole approach has recently been given by Kaplan\cite{Kaplan}.

Including the two lowest order interactions with coupling constants $C_0$ and $C_2$, the 
effective Lagrangian for the nucleon field $N^T = (p,n)$ with mass $M$ can be written as
\beq  
     {\cal L}_0 &=& N^\dagger\left(i\del_t + {\nabla^2\over 2M}\right)N
              - C_0(N^T{\bg\Pi}N)\cdot(N^T{\bg\Pi}N)^\dagger  \nn \\
    &+& {1\over 2}C_2\left\{(N^T\anabla^2{\bg\Pi}N)\cdot(N^T{\bg\Pi}N)^\dagger + h.c.\right\}
                                                                 \label{Leff}       
\eeq
where the operator $\anabla = (\overrightarrow{\nabla} - \overleftarrow{\nabla})/2$.
The projection operators $\Pi_i$ enforce the correct spin and 
isospin quantum numbers in the channels under investigation. More specifically, for 
spin-singlet interactions $\Pi_i = \sigma_2\tau_2\tau_i/\sqrt{8}$ while for spin-triplet 
interactions $\Pi_i = \sigma_2\sigma_i\tau_2/\sqrt{8}$. This theory is now valid below 
an upper momentum $\Lambda$ which will be the physical cutoff when the theory is regularized 
that way. For momenta much smaller than the pion mass, we can consider the pion field integrated
out and all its effects soaked up in the two coupling constants $C_0$ and $C_2$. Then the
value of the cutoff $\Lambda$ will be set by the pion mass $m_\pi$. In this momentum range 
all the main properties of few-nucleon systems should then in principle be given by the above
Lagrangian. More accurate results will follow from higher order operators in this 
field-theoretic description\cite{CRS}.

From calculations of proton-neutron elastic scattering using the PDS regularization 
scheme\cite{KSW_1}\cite{KSW_2} or the equivalent OS scheme\cite{OS}, the {\it a priori} 
unknown coupling constants $C_0$ and $C_2$ can be determined in terms of experimental 
quantities. In the spin-triplet channel the deuteron will appear as a bound state and 
the corresponding lowest order renormalized coupling constant is
\beq
     C_0^d(\mu) = {4\pi\over M} {1\over\gamma - \mu}                      \label{C0d}
\eeq
where $\gamma = 45.7$ MeV is the momentum in the deuteron bound state\cite{KSW_2}\cite{CRS}. 
The renormalization 
mass $\mu$ can be chosen freely in the interval $\gamma < \mu \le m_\pi$ and physical results
should be independent of its precise value. Similarly, for the dimension-eight coupling 
constant  one finds
\beq
     C_2^d(\mu) = {4\pi\over M} \left({1\over\gamma - \mu}\right)^2{\rho_d\over 2} \label{C2d}
\eeq
where $\rho_d = 1.76$ fm is the spin-triplet $pn$ effective range scattering parameter 
evaluated at the deuteron pole. Together
with the similar coupling constants in the spin-singlet channel, many properties of
the proton-neutron system can be calculated. In particular, results for proton-neutron
radiative capture $n + p \ra d + \gamma$ have recently been obtained\cite{CRS}\cite{SSW} 
which from the hadronic point of view is very similar to proton-proton fusion $p + p \ra
d + e^+ + \nu_e$.

The coupling constants  in the $pp$ channel can also be matched to experimental scattering 
data at low energies. However, in this energy range the Coulomb effects become important 
and must be separated out. This has recently been done within the framework of the same 
effective theory\cite{KR_pp} and also within standard quantum mechanics\cite{BH}. The leading
order coupling constant in the spin-singlet $pp$ channel can now be written as
\beq
     C_0^p(\mu) = {4\pi\over M} {1\over 1/a(\mu) - \mu}                       \label{C0p}
\eeq     
where $a$ would be the strong scattering length when there were no Coulomb interactions.
But these turn it into a $\mu$-dependent quantity which can be determined from the measured
scattering length $a_p = - 7.82$ fm using 
\beq
    {1\over a(\mu)} = {1\over a_p}  +  \alpha M
    \left[\ln{\mu\sqrt{\pi}\over\alpha M} + 1 - {3\over 2}C_E\right]             \label{Capp}
\eeq
where $\alpha$ is the fine-structure constant and $C_E = 0.5772\ldots$ is Euler's constant. 
Since the proton mass is so much heavier than $1/a_p$, the Coulomb correction is seen to be
surprisingly large. This has been known for a long time and was previously expressed by the
corresponding Jackson-Blatt relation of the same form and obtained from potential 
models\cite{JB}. The higher order coupling constant $C_2^p$ in this channel will have the 
same form as $C_2^d$ in (\ref{C2d}) but with $\gamma$ replaced with $1/a(\mu)$ and $\rho_d$ 
with the proton-proton effective range $\rho_p = 2.79$ fm. We have shown that it is not 
affected by Coulomb corrections to this order in the effective theory. However,
the $C_2^p$ coupling gives an important contribution to the scattering length (\ref{Capp})
which picks up an additional term $-\mu\rho_p/2$ in the parenthesis\cite{KR_pp}.

Including only the leading order couplings $C_0^{p,d}$, we have recently calculated the
rate for proton-proton fusion $p + p \ra d + e^+ + \nu_e$ as it takes place in the 
Sun\cite{KR_fus}. Since the initial energy $E = p^2/M$ is then so low, it can be taken to zero.
Thus also the momenta of the final state leptons can be ignored. Stating the result
in terms of the standard reduced matrix element $\Lambda(E)$ which is 
dimensionless\cite{ES}\cite{BM}, we obtained the result
\beq
    \Lambda_0(0) = e^\chi - 2\alpha M a_p\,I(\chi)                           \label{Lambda0}
\eeq
where the parameter $\chi = \alpha M/\gamma$. Here we have introduced the function
\beq
    I(\chi) = {1\over\chi} - e^\chi E_1(\chi)                              \label{Ichi}
\eeq
where $E_1(\chi)$ is the exponential integral function. This is in full agreement with
the corresponding result obtained in the zero-range approximation of nuclear potential
models\cite{BM}. With the above values for the different parameters, we 
have $\chi = 0.15$ and $I(0.15) = 4.96$. The matrix element then becomes $\Lambda_0(0) 
= 2.51$ so that in the rate we have $\Lambda_0^2(0) = 6.30$. This is 10\% below the value 
$\Lambda_{tot}^2(0) = 7.0 \pm 0.05$ obtained in the most complete calculations including 
higher order effects\cite{BK_1}\cite{nucl}.

When we now include the effects of the next order coupling $C_2$ to the fusion process, 
it gives rise to the Feynman diagrams shown in Fig.1. In the
chain of bubbles connected by $C_0$ interactions, the two protons interact via the
Coulomb potential $V_C(r) = \alpha/r$. In a single bubble the particles propagate
from zero separation and back to zero separation. The contribution of a single bubble in
the chains is thus $J_0(p) = G_C(E;0,0)$ where $G_C(E;\br,\br')$ is the Coulomb propagator
in coordinate space of the two protons with energy $E$. We have found it most 
convenient to calculate the value of the bubble quantity $J_0(p)$ in momentum space. With
$\int_\bq \equiv \int\!d^3 q/ (2\pi)^3$ it is
\beq
     J_0(p) = \int_{\bq}\int_{\bq'} G_C(E;\bq,\bq')
\eeq
when expressed in terms of the Fourier-transformed Coulomb propagator $G_C(E;\bq,\bq')$.
This integral can now be done using either dimensional regularization in the PDS scheme 
or a standard momentum cutoff to regularize the ultraviolet divergence it 
contains\cite{KR_pp}\cite{BH}. The result has been used to obtain the Coulomb-corrected
scattering length (\ref{Capp}) and the leading order fusion result (\ref{Lambda0}). 

\begin{figure}[ht]
 \begin{center}
  \epsfig{figure=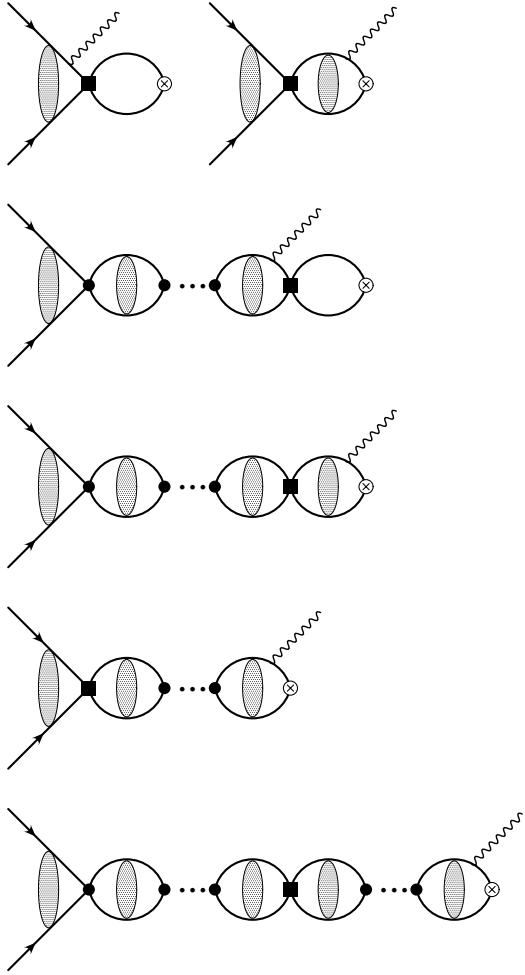,height=129mm}
 \end{center}
 \vspace{-4mm}            
 \caption{\small Feynman diagrams contributing to the fusion process to
first order in the derivative coupling $C_2$ denoted by black square. The
wiggly line indicates the weak current while the cross circle denotes the
action of the deuteron interpolating field.}
 \label{fig.1}
\end{figure}

In some of the diagrams in Fig.1 the derivative operator $\anabla^2$ acts on such 
Coulomb-dressed bubbles and gives rise to more divergent integrals. One of the simplest
is
\beq
     J_2(p) =  \int_{\bq}\int_{\bq'}\bq^2 G_C(E;\bq,\bq')               \label{J2}
\eeq
Using a method based upon the the Fourier-transformed Coulomb wavefunctions, we have
obtained a finite result for this integral\cite{KR_pp}. However, the method lacks a firm
foundation and cannot easily be applied to other integrals of a more general form. We will
therefore here use a more direct and simpler method based upon the functional equation satisfied
by the Coulomb Green's function\footnote{This new method arose out of a discussion with Peter
Lepage to whom we are much thankful.}. Introducing the free propagator
\beq
       G_0(E;\bq,\bq') =  {M\over \bp^2 - \bq^2 + i\e}(2\pi)^3\delta(\bq - \bq')
\eeq
the functional equation is $G_C = G_0 + G_0V_CG_C$. The integral (\ref{J2}) can then be written
as
\be
     J_2(p) &=& p^2 J_0(p) +  \int_{\bq}\int_{\bq'}
                    (\bq^2 - \bp^2) G_0(E;\bq,\bq') \\
           &+&   \int_{\bq}\int_{\bq'} \int_{\bk}\int_{\bk'} 
                 (\bq^2 - \bp^2)G_0(E;\bq,\bk) V_C(\bk,\bk') G_C(E;\bk',\bq')
\ee                                                                          
where $V_C(\bk,\bk')= 4\pi\alpha/(\bk - \bk')^2$ is the Fourier transform of the Coulomb 
potential. The first integral
is now zero by dimensional regularization. Integrating over $\bk$ in the second, we see that
the first two factors just give $-M$. In the integration of $V_C(\bq,\bk')$ over $\bq$, we can
now shift the integration variable  $\bq \ra \bq - \bk'$ and use
\beq
        \int_\bq {4\pi\alpha\over \bq^2} = \alpha\mu                      \label{coul}
\eeq
This integral is also zero with dimensional regularization, but it contains a PDS pole in
$d=2$ space dimensions which gives the non-zero contribution. The remaining integrations 
over $\bq'$ and $\bk'$ give then finally simply $J_0(p)$. Collecting all the factors, we 
thus obtain the simple result
\beq
     J_2(p) = (p^2 - \alpha\mu M)J_0(p)
\eeq
This agrees with the result using our previous method except for terms which are smaller
by factors of the order of $\alpha M/\mu$. These can in practice be neglected since
we always take the renormalization point $\mu \gg \alpha M$.

Using the same method to evaluate other similar integrals arising from the Feynman diagrams in
Fig.1, we can then simplify and gather all the contributions into a modified reduced matrix
element. A part of this new contribution goes into the next-to-leading order scattering
length $a_p$ in the lowest order result (\ref{Lambda0}). The remaining terms simplify to
\beq
     \Lambda_2(0) =  C_2^d{M\gamma^2\over 4\pi}(\mu - \gamma) \Lambda_0(0)
                  -  a_p\gamma^2(\mu - \gamma) {C_2^p + C_2^d\over 2C_0^p}
\eeq
However, there is also a wavefunction renormalization constant $\sqrt{\Sigma'}$ which to this 
order will modify the lowest order result (\ref{Lambda0}). It will enter in all calculations
involving the bound state deuteron and has been calculated by Kaplan, Savage and 
Wise\cite{KSW_2}. In this version of the effective theory without pions it is given by
\beq
    \Sigma' = 1 -   C_2^d{M\gamma\over 2\pi}(\mu - \gamma)(\mu - 2\gamma)
\eeq
The full result for the reduced matrix element in next-to-leading order is thus
\beq
    \Lambda_{NLO} = {\Lambda_0\over\sqrt{\Sigma'}} + \Lambda_2
\eeq
Expanding this now to first order in $C_2^d$ with the renormalized value (\ref{C2d}), 
we get the final result
\beq
    \Lambda_{NLO}(0) =   \Lambda_0(0)\left(1 + {1\over 2}\gamma\rho_d\right)
                  - a_p\gamma^2(\mu - \gamma) {C_2^p + C_2^d\over 2C_0^p}    \label{NLO}
\eeq
The last term is seen to be dependent in general on the renormalization mass $\mu$. From a 
physical point of view, the result should be independent of this arbitrary parameter. 
What makes
this possible here, is the presence of a new, local interaction which comes in as a counterterm
at this order of perturbation theory. It will be $\mu$-dependent in such a way as to make the
overall result independent of $\mu$. A very similar situation arises in the process 
$n + p \ra d + \gamma$ where such a counterterm also is present\cite{CRS}\cite{SSW}. 
The {\it a priori} magnitudes of these counterterms are determined by physics on scales 
shorter than included in the effective theory. An absolute prediction of the proton-proton 
fusion rate is thus not possible at this next-to-leading order as long as this counterterm
is not determined by other means.

Here we will instead  compare our result (\ref{NLO}) with the corresponding result from
potential models. When we take $\mu \gg \gamma$, the dependence of the result on this
arbitrary renormalization mass becomes negligible and we find
\beq
    \Lambda_{NLO}(0)_{\mu\gg\gamma} =   \Lambda_0(0)\left(1 + {1\over 2}\gamma\rho_d\right)
                     + {1\over 4}a_p\gamma^2(\rho_p + \rho_d)              \label{NLOX}
\eeq
With the previous values of the different parameters, we obtain for the reduced matrix
element the value $\Lambda(0) = 2.54$ which is just a 1.4\% addition to the leading
order result. This is surprisingly small, but results from an almost total
cancellation between the two effective-range corrections in (\ref{NLOX}).
In the full rate, it corresponds to a value which is 
8\% below the accepted value from nuclear potential models\cite{BK_1}\cite{nucl}.
However,
the structure of our result is very similar when compared to what one obtains in the
corresponding effective range approximation\cite{BM}. The last term in (\ref{NLOX}) is then
exactly the same, while in the first term the factor $(1 + {1\over 2}\gamma\rho_d)$ is
replaced by $(1 - \gamma\rho_d)^{-1/2}$ which in nuclear models is the normalization factor
of the deuteron wavefunction. To lowest order in the deuteron effective range
parameter $\rho_d$ this is then just the same. But $\gamma\rho_d = 0.41$ is really not a
small expansion parameter and higher order terms in the expansion of the square root give a
sizeable contribution. These can only be reproduced in the effective theory when including
even higher order interactions. At the same time this will then also bring in new and unknown
counterterms. It thus seems difficult for this effective theory to compete with the more
accurate results obtained from potential models where the deuteron effective range corrections 
are included to all orders.

A more detailed presentation of the calculation and discussion of our results including the 
presence of the counterterm, will be presented elsewhere.

We want to thank the organizers of the workshop on ``Nuclear Physics with Effective Field 
Theory'' at the Institute of Nuclear Theory during which this work came to conclusion. 
In addition, we are grateful to the Department of Physics and the INT for generous support 
and hospitality. Xinwei Kong is supported by the Research Council of Norway.

\end{document}